\documentclass[10pt]{iopart}

\usepackage[T1]{fontenc}
\usepackage[utf8]{inputenc}
\usepackage{graphicx}
\usepackage{tabularx}
\usepackage{color, soul}
\usepackage{textcomp}
\usepackage{enumerate}
\usepackage{todonotes}
\usepackage[draft]{hyperref}
\usepackage{comment}
\usepackage{subcaption}
\usepackage{url}
\usepackage[normalem]{ulem}
\usepackage{makecell}
\usepackage{graphbox}
\usepackage{stfloats}
\usepackage{booktabs} % For formal tables
\usepackage{multirow}
\usepackage{cite}
\usepackage{amssymb}

\providecommand{\RV}[1]{#1}

\begin{document}

\title[Investigation of Stimulated Brillouin Scattering Driven by Broadband Lasers in High-Z Plasmas]{Investigation of Stimulated Brillouin Scattering Driven by Broadband Lasers in High-Z Plasmas} 

\author{Xiaoran Li$^{1}$, Jie Qiu$^{1}$, Liang Hao$^{1}$\footnote{Corresponding author: hao\_liang@iapcm.ac.cn}, Shiyang Zou$^{1}$}

\address{
$^1$ Institute of Applied Physics and Computational Mathematics, Beijing 100094, People’s Republic of China}

%\ead{hao\_liang@iapcm.ac.cn}

\vspace{10pt}
\begin{indented}
        \item[]\today
\end{indented}

\begin{abstract}
The evolution of stimulated Brillouin scattering (SBS) driven by broadband lasers in high-Z plasmas is investigated using one-dimensional collisional particle-in-cell simulations.
The temporal incoherence of broadband lasers modulates the pump intensity, generating stochastic intensity pulses that intermittently drive SBS.
The shortened coherence time weakens the three-wave coupling and continuously reduces the temporal growth rate, while the saturated reflectivity remains nearly unchanged until the bandwidth exceeds a critical threshold.
Simulations with varying laser intensities and bandwidths reveal a consistent scaling behavior, indicating that effective suppression occurs only when the laser bandwidth exceeds the temporal  growth rate of SBS by several tens of times.
Comparative simulations in Au and AuB plasmas exhibit similar suppression trends, with AuB showing reduced SBS growth rate and reflectivity, and the onset of suppression occurring at a lower bandwidth.
These findings elucidate the coupled dependence of SBS mitigation on bandwidth and laser intensity in high-Z plasmas, offering useful guidance for optimizing broadband laser designs in inertial confinement fusion.
\end{abstract}

\begin{indented}
\item[] \textbf{Keywords}: laser-driven inertial confinement fusion, laser plasma instability, stimulated Brillouin scattering, broadband lasers, high-Z plasmas
\end{indented}

% Uncomment for two-column
\ioptwocol

\section{Introduction}

\RV{Laser-driven inertial confinement fusion (ICF) is a promising approach to controlled thermonuclear fusion, which may provide a viable route toward future clean energy sources~\cite{hurricane2023}.}
The energy coupling between the incident laser and plasmas is strongly affected by laser plasma instabilities (LPI)~\cite{michel2023}, which can lead to significant laser energy loss and the generation of super-thermal electrons, thereby reducing the efficiency of ICF~\cite{montgomery2016,kline2019}.
LPIs have long been identified as a critical challenge in various approaches to laser-driven ICF~\cite{betti2016}.
Among various LPI mitigation strategies, broadband lasers have recently attracted growing attention~\cite{zhang_nonlinear_2025,liu_parametric_2023,bates_suppressing_2023,yin_mitigation_2025,bates_mitigation_2018,blackman_impact_2024} and emerged as a promising approach, facilitated by the rapid development of broadband laser facilities~\cite{gao2020,dorrer2021}.

Broadband laser sources, by introducing frequency spread and temporal
incoherence, are expected to disrupt the phase matching conditions required for
parametric instabilities, thus suppressing their growth.
Traditional theoretical models suggest that when the laser bandwidth $\Delta \omega$ is obviously larger than the linear temporal growth rate $\gamma$ of a given instability, the amplification of that instability can be effectively suppressed~\cite{thomson1974}.
Recent simulation studies have revealed a
more nuanced picture, particularly for stimulated Raman scattering (SRS) and
two-plasmon decay (TPD).
For instance, Zhao \textit{et al}~\cite{zhao_stimulated_2017} reported that while broadband lasers can increase the time duration for nonlinear saturation of SRS, it does not significantly alter the saturation level.
Liu \textit{et al}~\cite{liu_non-linear_2022} further demonstrated that when bandwidth is insufficient, intensity fluctuations of broadband lasers can even enhance SRS reflectivity under certain conditions.
Similarly, Yao \textit{et al}~\cite{yao_anomalous_2024} observed that this phenomenon extends to TPD, that TPD modes' energy and resulting hot electron energy oscillate at the modulation frequency when the frequency is at the same order as the TPD growth rate, highlighting the complex interplay between broadband lasers and LPIs.
	
In contrast to extensive investigations on SRS~\cite{zhang_nonlinear_2025, liu_fluid_2025, yin_mitigation_2025} and TPD~\cite{yao_anomalous_2024, yao_resonance_2025, liu_parametric_2023, bates_suppressing_2023}, studies on how broadband lasers influence stimulated Brillouin scattering (SBS) remain comparatively limited.  
Recent experiments conducted on the Kunwu laser facility have demonstrated that a broadband laser with a 0.6\% bandwidth can reduce the SBS reflectivity in CH plasmas compared to a monochromatic laser~\cite{kang_effects_2025, lei_reduction_2024, wang_backward_2024}.  
However, both these experimental results and related computational studies~\cite{zhao_mitigation_2024} were restricted to low-Z and mid-Z plasmas, without considering high-Z targets.  
Since the evolution of laser plasma instabilities is highly sensitive to plasma parameters~\cite{qiu2021b}, it remains uncertain whether the mitigation effects observed in low-Z systems can be extended to high-Z conditions.  
Given that SBS is generally more pronounced in high-Z plasmas due to their lower ion-acoustic velocities and reduced Landau damping, the influence of broadband lasers under such conditions warrants further systematic investigation.

In this work, we investigate how broadband lasers influence the evolution of SBS in Au plasmas, which serve as a representative high-Z material commonly used as hohlraum walls in indirect-drive ICF.
The simulation model and the broadband laser characteristics are introduced in section~\ref{sec:model}.
Section~\ref{sec:m_vs_b} presents a comparison between SBS driven by broadband and monochromatic lasers, followed by section~\ref{sec:bandwidth}, which analyzes how varying bandwidth affects the temporal evolution and saturation behavior.
The dependence of these effects on laser intensity is then examined in section~\ref{sec:intensity}.
Broadband laser driven SBS in AuB plasmas is explored in section~\ref{sec:AuB}, providing further insight into the influence of ion composition and bandwidth on SBS suppression.
Finally, the main conclusions are summarized in section~\ref{sec:conclusion}.

\section{Simulation Model and Parameters}
\label{sec:model}

To investigate the influence of broadband lasers on the evolution of SBS in Au and AuB plasmas, a series of one-dimensional three-velocity (1D3V) particle-in-cell (PIC) simulations are performed using the open-source code \textsc{EPOCH}~\cite{arber2015}.  
The simulation domain has a total length of $1000\,c/\omega_0$, where $\omega_0$ is the frequency of the incident laser with a wavelength of $\lambda_0 = 0.351\,\mu$m.  
The plasma is initialized as spatially uniform with an electron density of $n_e = 0.3\,n_c$, where $n_c = 8.9\times10^{21}\,\mathrm{cm^{-3}}$ is the critical density corresponding to a 0.351\,$\mu$m laser.  
The ion densities are determined by charge neutrality according to the ion charge states.  
For pure Au plasmas, the ion density is $n_{\mathrm{Au}} = n_e / 50$, while for AuB plasmas, both Au and B ions have densities of $n_e / 55$.  
The electron and ion temperatures are uniform throughout the domain, with $T_e = 3000$\,eV and $T_i = 1000$\,eV, respectively.
The plasma region of $1000\,c/\omega_0$ is discretized into 10,000 cells, giving a spatial resolution of $\Delta x = 0.1\,c/\omega_0$.  
Two convolutional perfectly matched layers~\cite{roden2000}, each consisting of 80 grids, are applied at both ends of the domain to absorb outgoing electromagnetic waves.  
%Open boundary conditions are used for all particles to prevent artificial reflections.  
The simulations employ 1000 macro-particles per cell for electrons and 100 macro-particles per cell for Au. The number of macro-particles for B ions is also 100 when considering AuB plasmas.
The total simulation time is approximately 30\,ps for all cases, ensuring a sufficient duration to capture both the linear growth and the nonlinear saturation of SBS.
Since this work focuses on high-Z plasmas, where collisional effects play a significant role, binary Coulomb collisions among all particle species are included.  
The collisions between electrons, ions, and between different ion species are modeled using the Nanbu Monte Carlo binary collision algorithm~\cite{nanbu1998}.

%\subsection{Laser Configuration and Broadband Modeling}

In numerical simulations, two primary approaches are commonly used to model broadband lasers: the frequency modulation method and the multi-frequency beamlet method.  
In the frequency modulation approach, the bandwidth effect is introduced by imposing a modulation on the fundamental frequency.  
In contrast, the multi-frequency beamlet method approximates a continuous power spectrum by superposing a large number of discrete frequency components.  
This beamlet based representation has been widely adopted in previous studies of broadband laser plasma interactions~\cite{liu_non-linear_2022, zhang_nonlinear_2025}, 
as it provides a simple yet accurate way to reproduce the essential temporal incoherence of broadband laser.  
Here we also adopt the multi-frequency beamlet method.

In the multi-frequency beamlet model, the incident laser electric field can be expressed as
\begin{equation}
	E(t) = \sum_{i=1}^{N} E_i \cos(\omega_i t + \phi_i),
\end{equation}
where $N$ is the total number of beamlets and $\phi_i$ denotes the random phase of the $i$th component, distributed in the range $[0, 2\pi]$.  
The beamlet frequencies $\omega_i$ are within the spectral interval centered at the fundamental frequency $\omega_0$ with a total bandwidth of $\Delta \omega$, i.e.,
$\omega_i \in \left[\omega_0 - \Delta \omega/2,\; \omega_0 + \Delta \omega/2 \right].$
The amplitude of each frequency component is determined by
$E_i = \sqrt{{h_{\omega}} I(\omega_i)},$
where $h_{\omega} = \Delta \omega / N$ represents the mean frequency spacing, and $I(\omega_i)$ is the power spectral density of the pump laser at frequency $\omega_i$.

In this study, the broadband laser is modeled using $N = 100$ beamlets. We adopt a flat-top power spectrum, where all beamlets have identical amplitudes.
The frequencies are uniformly distributed within the prescribed bandwidth, and the phases are randomly assigned in $[0, 2\pi]$.

\begin{figure*}[htb]
	\centering
	\includegraphics[width=0.85\textwidth]{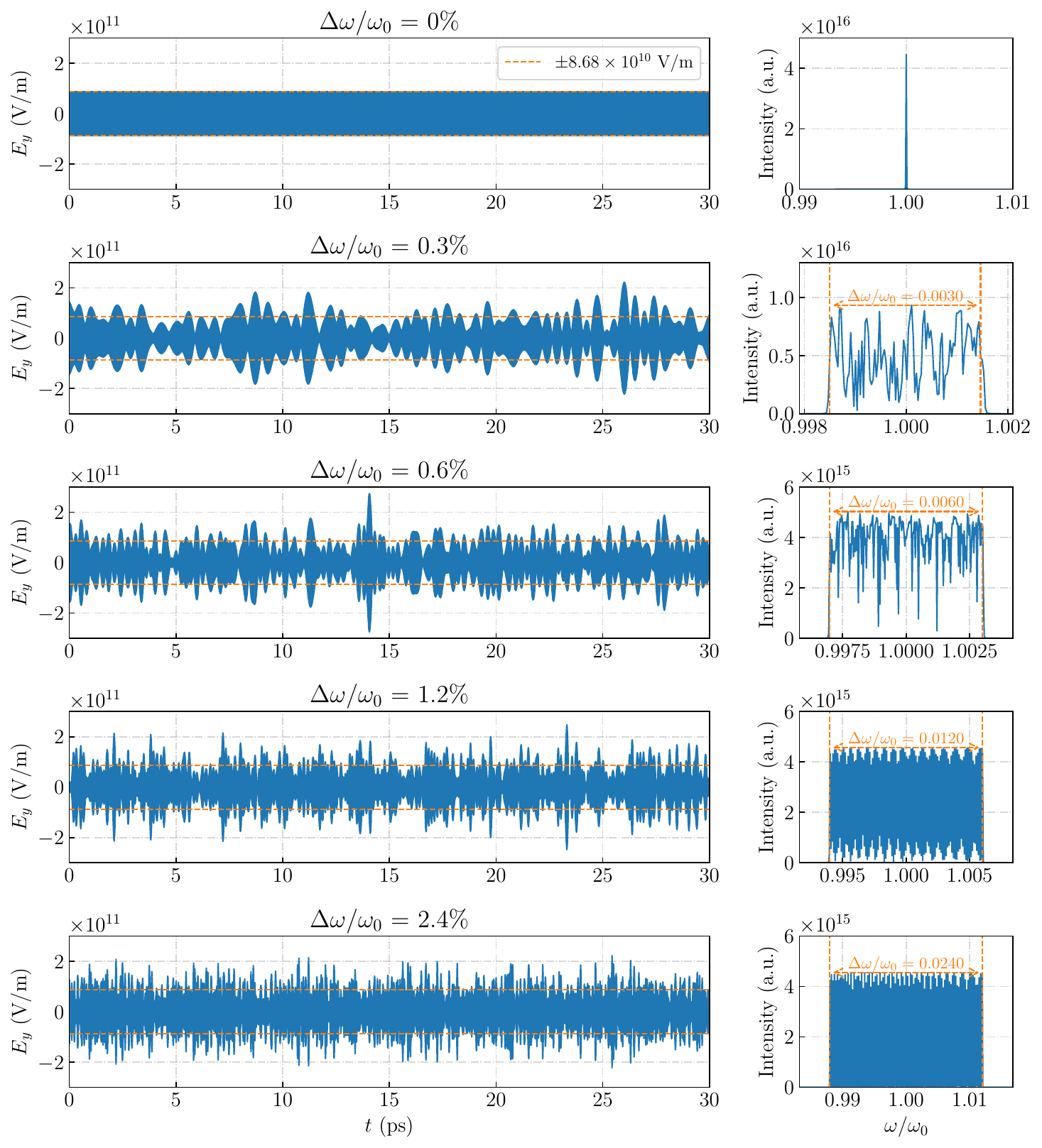}
	\caption{Temporal and spectral characteristics of the incident lasers with different bandwidths.
		The left panels show the time-dependent electric field profiles for bandwidths of 0 \%, 0.3\%, 0.6\%, 1.2\%, and 2.4\% relative to the central frequency.
		The right panels present the corresponding power spectra obtained by Fourier transform, confirming that the broadband laser fields used in the simulations reproduce the designed bandwidth.}
	\label{fig:lasers}
\end{figure*}

To systematically examine the bandwidth effects, five types of incident lasers are considered: a monochromatic laser and broadband lasers with bandwidths of 0.3\%, 0.6\%, 1.2\%, and 2.4\% relative to the central frequency.  
Figure~\ref{fig:lasers} presents the temporal waveforms and their corresponding power spectra obtained via the Fourier transform.  
For the monochromatic case ($\Delta \omega / \omega_0 = 0$), the electric field remains temporally uniform, corresponding to a single sharp spectral peak at $\omega = \omega_0$.  
When a finite bandwidth is introduced, the temporal waveform exhibits rapid fluctuations in amplitude, and the irregularity of the carrier oscillation period implies variations in the instantaneous phase caused by the random interference among multiple frequency components.  
These fluctuations become progressively stronger as the bandwidth increases, producing pronounced temporal intensity modulations.  
At a bandwidth of $\Delta \omega / \omega_0 = 0.3\%$, the field still retains partial temporal coherence, and the envelope shows slow oscillations.  
For $\Delta \omega / \omega_0 = 0.6\%$ and $1.2\%$, the field exhibits strong amplitude fluctuations on the picosecond scale, and the instantaneous intensity varies over a wide dynamic range.  
When the bandwidth increases to $2.4\%$, the temporal coherence is almost completely destroyed, and the intensity profile becomes highly stochastic, with only small-scale modulations around the mean level.  
This behavior reflects the inverse relationship between the bandwidth and the laser coherence time, $\tau_c \sim 1 / \Delta \omega$, 
such that a larger bandwidth corresponds to a shorter coherence time and hence stronger temporal incoherence of the pump field~\cite{mandel1962}.

In the frequency domain, the spectra for the broadband cases confirm that the implemented laser fields in the simulations accurately reproduce the designed bandwidths.  
It is noted that the spectrum for the $\Delta \omega / \omega_0 = 0.3\%$ case appears slightly nonuniform, with small deviations from the expected spectral shape.  
This discrepancy arises from the extremely narrow bandwidth, where the frequency spacing between adjacent components becomes comparable to the FFT resolution.  
As a result, the reconstructed spectrum from the finite-time sampled simulation data is more susceptible to numerical noise, leading to minor fluctuations around the designed spectral profile.  
Nevertheless, the overall bandwidths of the broadband laser fields remain consistent with the intended design.  
%These results confirm that increasing the bandwidth shortens the laser coherence time, a key factor for mitigating the growth of parametric instabilities.  

In the subsequent investigation of the influence of laser intensity on SBS under broadband conditions, the laser intensity is varied from $5\times10^{14}$ to $4\times10^{15}\,\mathrm{W/cm^2}$.  
For each bandwidth, only the field amplitude is adjusted to achieve the desired intensity, while the temporal waveform and phase structure of the laser remain unchanged.  
In all simulations, the laser is injected from the left boundary of the domain and propagates along the $+x$ direction toward the plasma target.

\section{Simulation Results}

\subsection{Comparison between SBS Driven by Monochromatic and Broadband Lasers}
\label{sec:m_vs_b}

\begin{figure}[htb]
	\centering
	\includegraphics[width=0.48\textwidth]{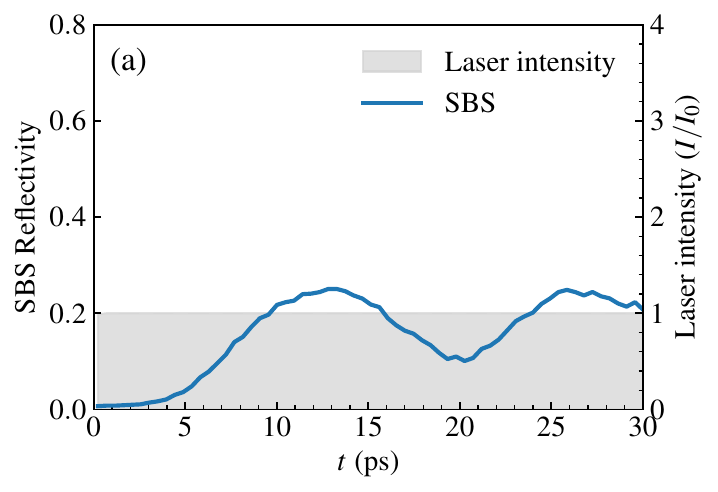}
	\includegraphics[width=0.48\textwidth]{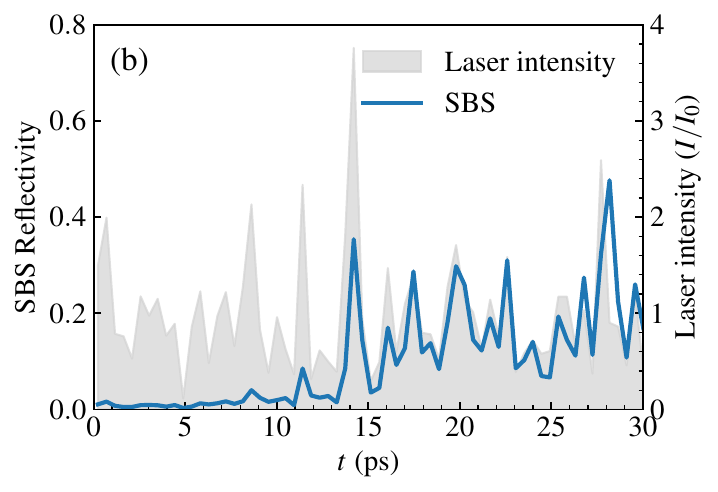}
	\caption{Comparison of the temporal evolution of SBS reflectivity and the corresponding laser intensity for 
		(a) a monochromatic laser and 
		(b) a broadband laser with $\Delta\omega / \omega_0 = 0.6\%$ in Au plasmas 
		at a laser intensity of 
		$I = 1\times10^{15}\,\mathrm{W/cm^2}$.  }
	\label{fig:narrow_vs_broad}
\end{figure}

To illustrate the fundamental differences between SBS driven by monochromatic and broadband lasers, two representative simulations are compared in figure~\ref{fig:narrow_vs_broad}.
The upper panel corresponds to the monochromatic laser case, while the lower panel shows the broadband laser with $\Delta\omega / \omega_0 = 0.6\%$.
Both simulations are performed at a laser intensity of $I = 1\times10^{15}\,\mathrm{W/cm^2}$ in Au plasmas.
The figure presents the temporal evolution of the incident laser intensity and the resulting SBS reflectivity.
For quantitative consistency, the SBS reflectivity is calculated from the backscattered field using a sliding-window FFT with a window length of $250\,\omega_0^{-1}$, and the laser intensity shown for comparison is averaged over the same temporal window.

In the monochromatic case, the incident intensity remains constant over time, and the SBS reflectivity exhibits a smooth and gradual rise before reaching a quasi-steady level.
Such behavior is typical of a coherent parametric process, where a single-frequency pump wave continuously drives a resonant ion-acoustic wave, leading to the exponential amplification of the backscattered light~\cite{li_langdon_2025}.

In contrast, when a finite bandwidth ($\Delta\omega / \omega_0 = 0.6\%$) is introduced, both the laser intensity and the SBS reflectivity display pronounced temporal fluctuations.
The instantaneous reflectivity rises and falls irregularly, forming a sequence of peaks and dips rather than a smooth growth curve.
This stochastic evolution originates from the rapidly varying interference pattern among multiple frequency components of the broadband laser.
The SBS fluctuations are strongly correlated with the temporal variations of the laser intensity on sub-picosecond timescales, indicating that the intensity oscillations of the broadband laser directly modulate the SBS process.
Notably, SBS driven by the broadband laser exhibits transient peaks exceeding those of the monochromatic case.
From an experimental perspective, this suggests that diagnostics and backscatter collection optics must be designed to withstand transient power surges, as instantaneous SBS reflectivity can momentarily exceed that driven by a monochromatic laser.

Compared with the monochromatic case, the broadband laser exhibits a clearly reduced early-time growth rate of SBS.
This reduction can be attributed to the shortened coherence time of the pump, where rapid phase decorrelation among the beamlet components intermittently disrupts the three-wave phase matching condition, thereby limiting the effective interaction length and weakening the net linear amplification, as suggested by reference~\cite{thomson1974}.
However, the time-averaged SBS reflectivity after saturation is comparable between the monochromatic laser ($R_{\mathrm{sat}} = 19.30\%$) and the $0.6\%$ broadband laser ($R_{\mathrm{sat}} = 19.38\%$).
Therefore, a bandwidth of $0.6\%$ primarily delays the SBS growth rather than reducing the final saturation level in Au plasmas at $I = 1 \times 10^{15}\,\mathrm{W/cm^2}$.
This behavior differs from previous experimental and simulation results in CH plasmas~\cite{kang_effects_2025, lei_reduction_2024, wang_backward_2024, zhao_mitigation_2024}, where a similar bandwidth was sufficient to significantly suppress SBS, indicating that the required mitigation bandwidth depends strongly on plasma compositions.
Motivated by this discrepancy, the following section systematically investigates how SBS evolves across a broader range of laser bandwidths.

\subsection{Bandwidth Dependence of SBS in Au Plasmas}
\label{sec:bandwidth}

\begin{figure}[htb]
	\centering
	\includegraphics[width=0.48\textwidth]{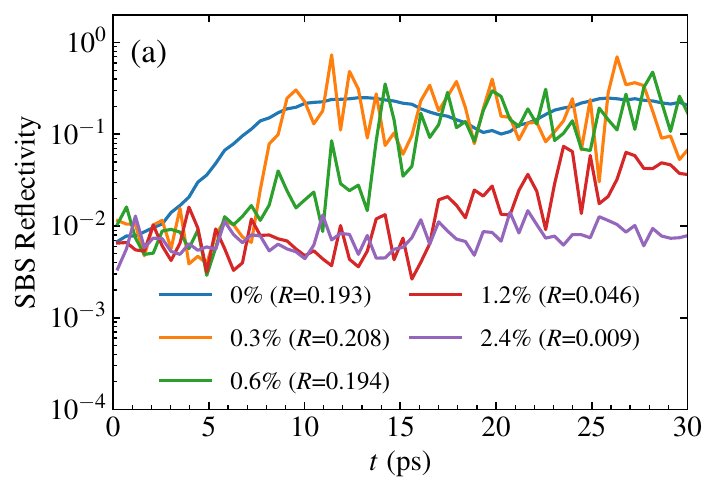}
	\includegraphics[width=0.48\textwidth]{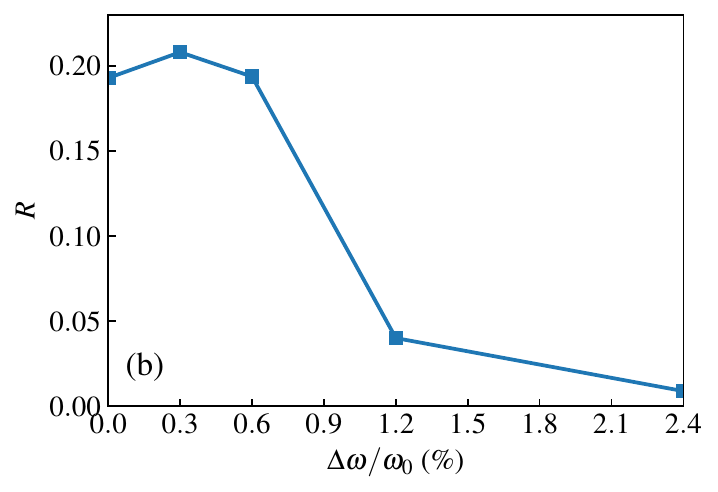}
	\includegraphics[width=0.48\textwidth]{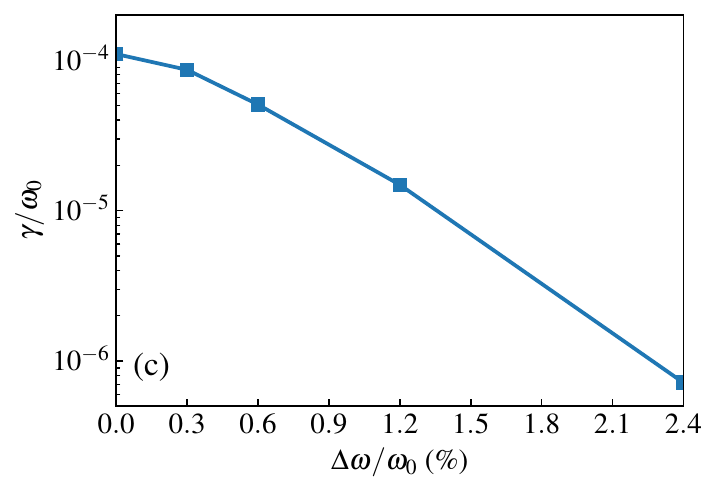}
	\caption{Dependence of SBS reflectivity and growth rate on laser bandwidth in Au plasmas 
		at a fixed laser intensity of 
		$I = 1\times10^{15}\,\mathrm{W/cm^2}$. 
		(a)~Temporal evolution of SBS reflectivity for different bandwidths. 
		(b)~Time-averaged SBS reflectivity $R$ and 
		(c)~temporal growth rate $\gamma/\omega_0$ as functions of laser bandwidth $\Delta\omega/\omega_0$. }
	\label{fig:bandwidth_sweep}
\end{figure}

To systematically investigate the influence of laser bandwidth on SBS in Au plasmas, 
a series of simulations is performed at a fixed laser intensity of 
$I = 1\times10^{15}\,\mathrm{W/cm^2}$, 
with the bandwidth varied from monochromatic to $2.4\%$.
Figure~\ref{fig:bandwidth_sweep} summarizes the results, 
showing the temporal evolution of SBS reflectivity for different bandwidths, 
together with the time-averaged reflectivity $R$ and the corresponding temporal growth rate $\gamma$ as functions of $\Delta\omega/\omega_0$.
For a consistent quantitative comparison, the average reflectivity $R$ is calculated after the onset of SBS saturation, 
defined as the moment when the instantaneous reflectivity first exceeds $10\%$.
This approach effectively eliminates the influence of different growth durations among cases on the averaged saturation level.
For simulations in which the reflectivity does not reach $10\%$ during the entire run, 
the average is instead taken over the last $10$\,ps of the simulation time.
\RV{The temporal growth rate $\gamma$ is obtained by fitting the early-time evolution of the computational reflectivity with an exponential function before saturation occurs. 
Physically, $\gamma$ represents the linear amplification rate of the backscattered light intensity, 
which characterizes the strength of three-wave coupling and the efficiency of energy transfer from the pump to the ion-acoustic wave and scattered light.}

The results reveal that the dependence of SBS on laser bandwidth exhibits two distinct trends: 
the temporal growth rate decreases monotonically with increasing bandwidth, 
whereas the time-averaged reflectivity remains nearly constant at small bandwidths and then drops abruptly beyond a critical threshold.

As shown in figure~\ref{fig:bandwidth_sweep}, when the bandwidth increases from monochromatic to $\Delta\omega/\omega_0 = 0.6\%$, 
the temporal growth rate of SBS steadily decreases from $\gamma/\omega_0 = 1.10\times10^{-4}$ to $5.08\times10^{-5}$. 
This continuous reduction indicates that the gradual loss of phase coherence between beamlets effectively weakens the three-wave coupling, thereby shortening the effective interaction time for acoustic wave amplification. 
However, in this range ($\Delta\omega/\omega_0 \le 0.6\%$), 
the mean SBS reflectivity does not show a clear decline. 
At a bandwidth of $\Delta\omega/\omega_0 = 0.3\%$, it even slightly exceeds the monochromatic value ($R= 20.8\%$ vs.\ $19.3\%$). 
This behavior indicates that a small amount of bandwidth introduces moderate phase decorrelation, 
which delays the growth onset but can locally enhance SBS through transient intensity peaks caused by temporal interference among beamlets. 
Such enhancement agrees with the trend observed in previous broadband SRS studies~\cite{liu_non-linear_2022}, 
where small bandwidths may amplify rather than suppress the instability due to intensity modulation. 
Therefore, in the small-bandwidth regime, the competing effects of phase decorrelation and local intensity bursts result in a near-balanced or slightly enhanced average SBS level.

Once the bandwidth exceeds $\sim1\%$, 
both the growth rate and average reflectivity experience a sharp decline. 
For $\Delta\omega/\omega_0 = 1.2\%$, the growth rate decreases by nearly an order of magnitude ($\gamma/\omega_0 = 1.49\times10^{-5}$), 
and the mean reflectivity drops to $R = 4.0\%$. 
At $\Delta\omega/\omega_0 = 2.4\%$, SBS is almost completely quenched, 
with $\gamma/\omega_0 = 7.18\times10^{-7}$ and $ R = 0.9\%$. 
In this regime, the pump coherence time becomes much shorter than the characteristic time required for ion-acoustic wave amplification, 
leading to an effective decorrelation between the pump and scattered light.

\RV{Overall, increasing the laser bandwidth effectively weakens SBS growth in Au plasmas, 
but the extent of suppression differs between the temporal growth rate and the reflectivity. 
The growth rate decreases steadily with increasing bandwidth, 
whereas the reflectivity is less sensitive and only shows a noticeable reduction 
once the bandwidth becomes sufficiently large.  
These observations suggest that the effectiveness of bandwidth mitigation 
is closely linked to the intrinsic SBS growth characteristics, 
which are expected to depend strongly on the laser intensity. 
Therefore, the following section explores how variations in laser intensity influence the bandwidth dependence of SBS.}

\subsection{Influence of Laser Intensity under Broadband Conditions}
\label{sec:intensity}

\begin{figure}[htb]
	\centering
	\includegraphics[width=0.48\textwidth]{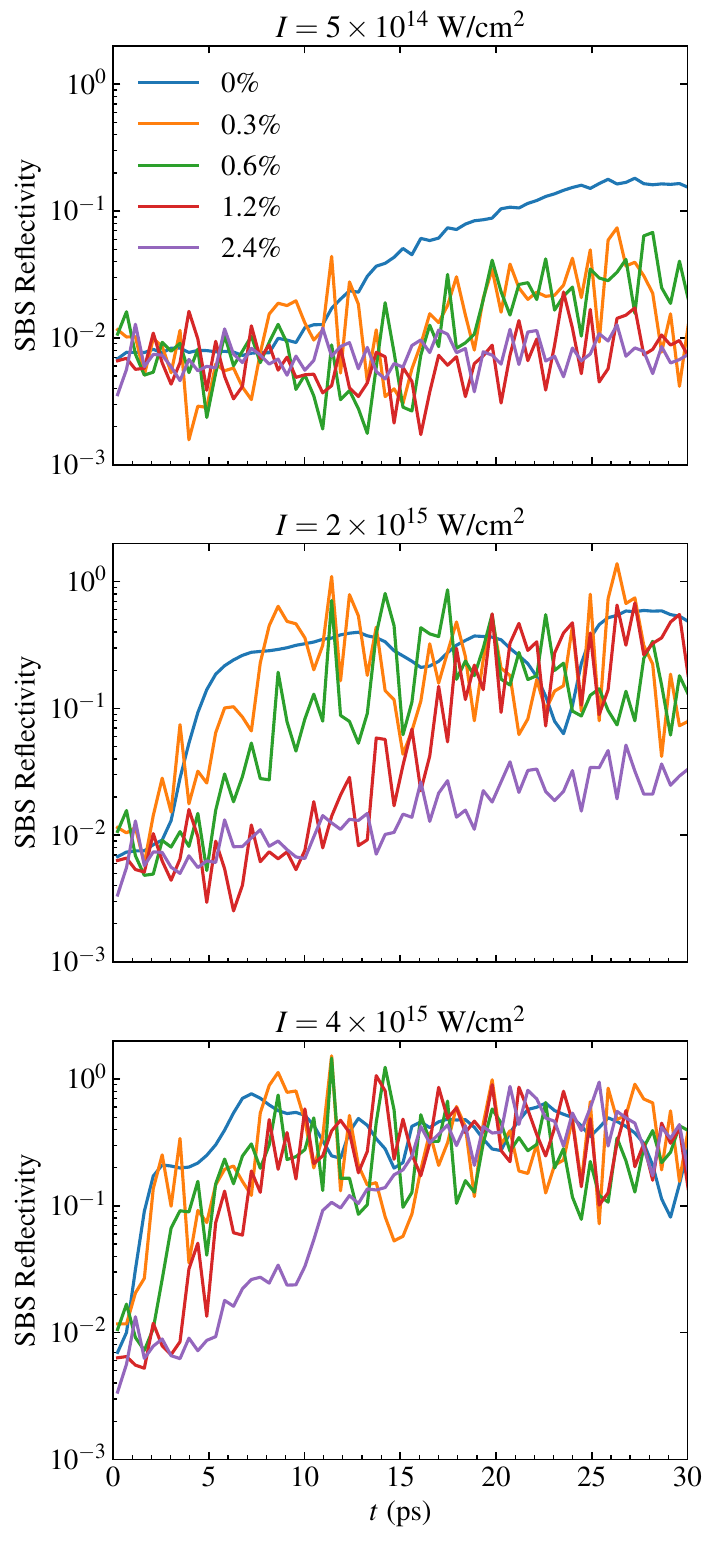}
	\caption{Temporal evolution of SBS reflectivity in Au plasmas for different laser bandwidths and incident laser intensities. }
	\label{fig:SBS_diffI}
\end{figure}

To further explore how the effectiveness of broadband mitigation depends on laser intensity, 
a series of simulations is performed for three additional intensities that are \(5\times10^{14}\), \(2\times10^{15}\) and \(4\times10^{15}\,\mathrm{W/cm^2}\), 
each with bandwidths of 0\%, 0.3\%, 0.6\%, 1.2\%, and 2.4\%.
Figure~\ref{fig:SBS_diffI} presents the temporal evolution of SBS reflectivity 
for different bandwidths at three representative laser intensities, 
providing an overview of how both the growth dynamics and saturation behavior vary with the pump strength.

For the monochromatic incident lasers, the SBS temporal growth rate \(\gamma\) increases almost linearly with laser intensity, 
taking values of \(4.18\times10^{-5}\omega_0\), \(1.10\times10^{-4}\omega_0\), \(2.69\times10^{-4}\omega_0\), and \(4.58\times10^{-4}\omega_0\) 
for $I$ = \(5\times10^{14}\), \(1\times10^{15}\), \(2\times10^{15}\) and \(4\times10^{15}\,\mathrm{W/cm^2}\), respectively.
This scaling is nearly proportional to the laser intensity and agrees with previous collisional studies of SBS in Au plasmas~\cite{li_langdon_2025}.  
When broadband spectra are introduced, the temporal growth rate is reduced in all cases, 
confirming that the finite coherence time weakens the linear amplification of SBS.  
However, the evolution of the saturated reflectivity shows a more intricate dependence on intensity.

At the lowest intensity of $5\times10^{14}\,\mathrm{W/cm^2}$, 
SBS is efficiently suppressed once the bandwidth exceeds $0.3\%$, 
with the mean reflectivity dropping from $R = 15\%$ (monochromatic) 
to $3\%$, indicating that even a modest reduction of coherence time can strongly weaken the instability.
At $1\times10^{15}\,\mathrm{W/cm^2}$, the suppression occurs only when the bandwidth reaches about $1.2\%$, 
where $R$ sharply decreases from $\sim19\%$ to $4\%$.  
At $2\times10^{15}\,\mathrm{W/cm^2}$, the reflectivity remains relatively high 
$R \approx 24\%\!-\!33\%$ for $\Delta\omega/\omega_0 \le 1.2\%$, 
but then drops sharply to $\sim 3\%$ at $\Delta\omega/\omega_0 = 2.4\%$. 
In contrast, at $4\times10^{15}\,\mathrm{W/cm^2}$ the reflectivity stays large 
$R \approx 32\%\!-\!39\%$ across all bandwidths considered up to $2.4\%$, 
indicating no appreciable suppression within this range.
These trends suggest that the critical bandwidth for suppression increases with intensity, 
as stronger pumps produce faster SBS growth requiring proportionally shorter coherence times to achieve comparable mitigation.

\begin{figure}[htb]
	\centering
	\includegraphics[width=0.48\textwidth]{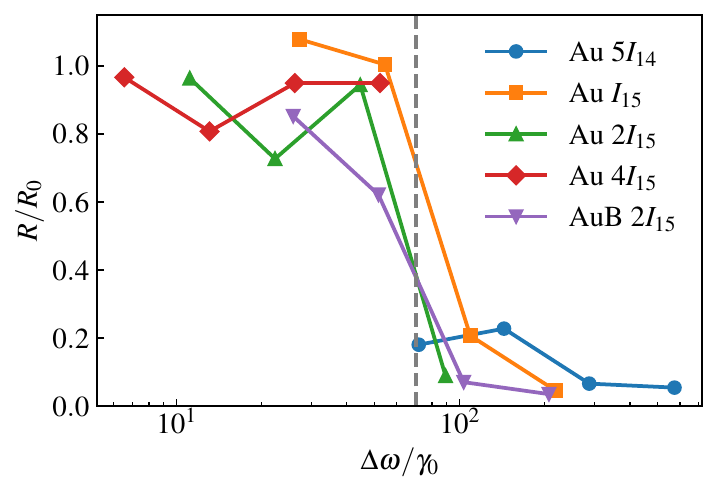}
	\caption{
		Dependence of normalized SBS reflectivity \(R/R_0\) 
		on the scaled bandwidth \(\Delta\omega/\gamma_0\) 
		for Au and AuB plasmas at different laser intensities.
		Here, \(R_0\) and \(\gamma_0\) denote the saturated reflectivity 
		and the temporal growth rate of SBS under monochromatic conditions, 
		respectively, with each intensity having its own corresponding \(R_0\) and \(\gamma_0\).
		The symbols \(5I_{14}\), \(I_{15}\), \(2I_{15}\), and \(4I_{15}\) correspond to incident laser intensity of  
		\(5\times10^{14}\), \(1\times10^{15}\), \(2\times10^{15}\), 
		and \(4\times10^{15}\,\mathrm{W/cm^2}\), respectively.
		%The vertical dashed line at \(\Delta\omega/\gamma_0 = 70\) marks the approximate onset 
		%of significant SBS suppression.
	}
	\label{fig:Rnorm_vs_scaledBW}
\end{figure}

To quantitatively compare the suppression behavior across different intensities, 
figure~\ref{fig:Rnorm_vs_scaledBW} shows the normalized SBS reflectivity \(R/R_0\) 
as a function of the scaled bandwidth \(\Delta\omega/\gamma_0\), 
where \(R_0\) and \(\gamma_0\) \RV{correspond to the saturated reflectivity and the fitted temporal growth rate obtained from the simulations of the monochromatic case at each intensity.}
The results reveal a consistent scaling across all intensities: 
SBS reflectivity remains nearly unchanged for $\Delta\omega/\gamma_0 < 70$, 
but decreases sharply once this threshold is exceeded.  
This demonstrates that effective suppression occurs only when the laser bandwidth 
is several tens of times larger than the SBS temporal growth rate, 
that is, when the pump coherence time becomes much shorter than the characteristic SBS amplification time.  
The interplay between these two timescales coherence-time and growth-time
governs the onset of broadband stabilization.

%\RV{In summary, broadband lasers consistently reduce the SBS growth rate at all intensities, 
%but the suppression threshold in reflectivity increases with the pump strength.  
%The present results indicate that significant mitigation requires the laser bandwidth to exceed 
%the SBS growth rate by more than an order of magnitude, 
%highlighting that merely shortening the coherence time moderately is insufficient to disrupt 
%the resonant three-wave coupling.}

\subsection{Extension to Multi-Ion Plasmas: SBS in AuB}
\label{sec:AuB}

To extend the analysis to multi-ion systems, we investigate the SBS behavior 
in an AuB plasma mixture at \(I = 2\times10^{15}\,\mathrm{W/cm^2}\) and with identical bandwidth settings as in the previous section. 
This comparison allows assessing how ion composition modifies the SBS growth and its response to laser bandwidth.

\begin{figure}[htb]
	\centering
	\includegraphics[width=0.48\textwidth]{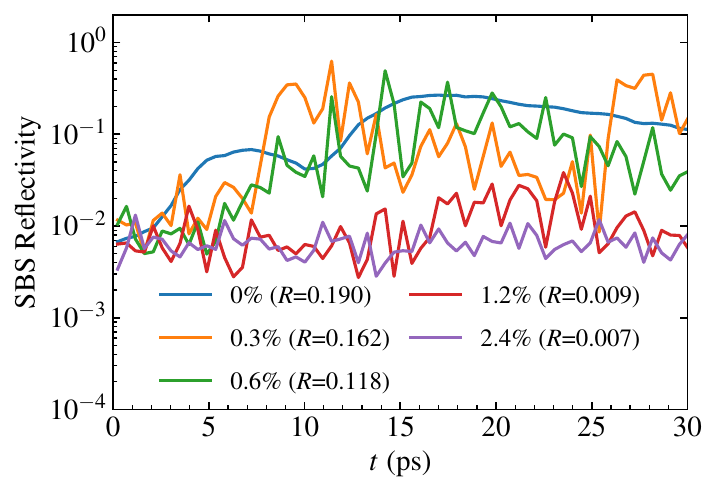}
	\caption{
		Temporal evolution of SBS reflectivity in AuB plasmas 
		at different laser bandwidths under \(I = 2\times10^{15}\,\mathrm{W/cm^2}\). 
		%Both the growth rate and average reflectivity are markedly lower than those in pure Au plasmas, indicating enhanced ion-acoustic damping due to the presence of light ions.
	}
	\label{fig:SBS_AuB}
\end{figure}

Figure~\ref{fig:SBS_AuB} shows the temporal evolution of SBS reflectivity 
for AuB plasmas at different bandwidths.
Compared with the pure Au case at the same intensity and the same bandwidth, 
both the temporal growth rate and the mean reflectivity are significantly reduced.  
For the monochromatic cases, the linear growth rate decreases from \(2.69\times10^{-4}\omega_0\) in Au 
to \(1.16\times10^{-4}\omega_0\) in AuB, 
and the saturated reflectivity $R$ decreases from 33\% to 19\%.
This reduction reflects the enhanced ion-acoustic damping introduced by the lighter B ions,
which lowers the effective gain of the three-wave coupling process~\cite{berger2023}.

In AuB plasmas, the dependence of SBS on laser bandwidth exhibits a trend similar to that in pure Au. 
The temporal growth rate \(\gamma\) continuously decreases as the bandwidth increases, 
indicating that the broadband-induced phase decorrelation weakens the parametric coupling in both systems.  
The saturated reflectivity shows a gradual decrease at small bandwidths 
and then drops to a very low level \(R \approx 1\%\) once the bandwidth reaches \(1.2\%\).  
In comparison, achieving a similar level of suppression in pure Au plasmas at the same laser intensity 
requires a bandwidth of about \(2.4\%\).  
Considering that both the monochromatic reflectivity and growth rate in AuB are already lower than those in Au, 
the earlier onset of suppression is consistent with the intuitive expectation that the reduced SBS reflectivity lowers the effective bandwidth required for stabilization.  

We further analyzed the scaling of the normalized reflectivity \(R/R_0\) 
with respect to the scaled bandwidth \(\Delta\omega/\gamma_0\) for the AuB cases, 
and plotted the results together with those for Au in figure~\ref{fig:Rnorm_vs_scaledBW}.  
The results show that, when normalized by their respective monochromatic parameters, 
the data points for AuB and Au collapse onto a nearly identical trend: 
the normalized reflectivity remains high at small \(\Delta\omega/\gamma_0\) 
and drops rapidly once the bandwidth becomes several tens of times larger than the growth rate.  
This indicates that the scaling of SBS suppression with the effective bandwidth 
is universal across single- and multi-ion high-Z plasmas, 
being primarily determined by the ratio of laser coherence time to the intrinsic SBS growth time.

%In summary, doping low-Z B ions into high-Z Au plasmas effectively reduces the SBS growth rate and saturation reflectivity by enhancing Landau damping of the ion-acoustic wave.
%When combined with broadband pump, these two suppression mechanisms can act synergistically to further mitigate SBS in high-Z hohlraum environments.

\section{\RV{Summary \& Discussion}}
\label{sec:conclusion}

In this work, we have systematically investigated the influence of broadband lasers on stimulated Brillouin scattering in high-Z plasmas using one-dimensional particle-in-cell simulations that include binary Coulomb collisions.  
The analysis focuses on Au and AuB plasmas as representative high-Z systems relevant to indirect-drive ICF, and clarifies how the laser bandwidth and intensity jointly determine the SBS growth rate, saturation behavior, and suppression threshold.

For monochromatic and broadband lasers, the temporal evolution of SBS exhibits markedly different characteristics.  
With a monochromatic incident laser, the SBS reflectivity grows smoothly and monotonically before reaching a quasi-steady saturation level, reflecting the coherent and continuous nature of the three-wave coupling.  
In contrast, when a finite bandwidth is introduced, the SBS signal shows pronounced temporal fluctuations that closely follow the instantaneous intensity variations of the broadband pump.  
These stochastic oscillations arise from the random interference among multiple frequency components, which intermittently modulates the effective driving field and leads to a burst-like SBS response.  

Introducing finite bandwidth reduces the SBS temporal growth rate at all intensities, confirming that temporal incoherence and phase decorrelation weaken the three-wave coupling responsible for amplification.  
However, the saturated reflectivity remains nearly unchanged until the bandwidth exceeds a critical threshold, beyond which SBS is strongly suppressed.  
At a fixed laser intensity of $1\times10^{15}\,\mathrm{W/cm^2}$, this transition occurs around $\Delta\omega/\omega_0 \simeq 1\%$, while for lower and higher intensities, the threshold shifts to smaller and larger bandwidths, respectively.  
When normalized by the monochromatic growth rate, all results collapse onto a consistent scaling law:  
the reflectivity remains nearly constant at small $\Delta\omega/\gamma_0$ and decreases sharply once the bandwidth becomes several tens of times larger than the temporal growth rate.  
\RV{In the present simulations, this transition occurs at $\Delta\omega/\gamma_0 \approx 70$, 
	indicating that effective SBS mitigation requires the pump coherence time to be much shorter than the characteristic SBS amplification time.  
	This value characterizes the current parameter regime and suggests that
	the effective suppression threshold lies within the range of several tens in $\Delta\omega/\gamma_0$ more generally, providing a useful reference for evaluating broadband mitigation of SBS in high-Z plasmas.} 
These findings also suggest that currently available experimental bandwidths~\cite{gao2020,dorrer2021} of $0.6\%\!-\!1\%$ 
may still be insufficient for strong SBS suppression in high-Z plasmas especially under high-intensity conditions.

Extending the analysis to AuB plasmas shows that the presence of light ions slightly enhances ion-acoustic damping and reduces both the SBS growth rate and the saturated reflectivity, leading to an earlier onset of suppression.  
When analyzed in terms of the normalized scaling $\Delta\omega/\gamma_0$, the results for Au and AuB exhibit nearly identical suppression trends, indicating that the underlying stabilization mechanism remains similar for these high-Z plasmas.  
The results suggest that broadband pumping and low-Z ion doping can work synergistically to mitigate SBS in high-Z hohlraum environments.

Overall, this study establishes a consistent physical picture of SBS driven by broadband lasers in high-Z plasmas, highlighting the coupled dependence of suppression efficiency on bandwidth and laser intensity.
\RV{The results provide useful physical insight and practical reference for designing broadband laser drivers to improve energy coupling and reduce backscatter losses in indirect-drive ICF experiments.}

\section*{Acknowledgments}
This work was supported by the National Natural Science Foundation of China (Grant Nos. 12505268, 12275032 and 12205021) and the China Postdoctoral Science Foundation (Grant No. 2024M764280).

\section*{Data availability statement}
The data cannot be made publicly available upon publication because the cost of preparing, depositing and hosting the data would be prohibitive within the terms of this research project. The data that support the findings of this study are available upon reasonable request from the authors.

\section*{References}

\bibliography{zotero-updating, wideband_clean}

\end{document}